\documentclass[MSNbibl,nameyear,dvips]{arxstspdf}
\usepackage{flushend}
\usepackage{stfloats}
\usepackage{graphicx}

\volume{26}
\issue{4}
\pubyear{2011}
\firstpage{604}
\lastpage{607}
\doi{10.1214/11-STS356C}
\referstodoi{10.1214/11-STS356}

\begin{document}
\begin{frontmatter}
\vspace*{6pt}
\title{Discussion of ``Multiple Testing for Exploratory Research'' by J. J. Goeman and
A. Solari}
\runtitle{Discussion}

\begin{aug}
\author[a]{\fnms{Peter H.} \snm{Westfall}\corref{}\ead[label=e1]{peter.westfall@ttu.edu}}
\runauthor{P. H. Westfall}

\affiliation{Texas Tech University}

\address[a]{Peter Westfall is Paul Whitfield Horn Professor of Statistics, Department of Information
Systems and Quantitative Sciences,  Texas Tech University, Lubbock, Texas 79409-2101, USA \printead{e1}.}

\end{aug}



\end{frontmatter}

\section{Initial Comments}

Closure-based multiple testing procedures for controlling the familywise error
rate (FWER) have been around for decades, but they have not been well
understood, and hence have been under-appreciated and under-utilized. Goeman
and Solari (GS) provide a service by highlighting important practical features
of closure. Using elegant notation for closure-based methods, they develop a
handy book-keeping tool for presenting additional results of closed testing
that are available when non-consonant testing methods are used, and they prove
its validity.

In their Figure 1, GS provide the confidence set $\tau(\{2,3\})\in\{0,1\}$,
where $\tau(\{2,3\})$ is the number of true nulls in the set $\{H_{2},H_{3}%
\}$. In doing so, GS highlight a not-so-well known fact about closure:
inferences for the additional $(2^{n}-1)-n$ composite hypotheses $H_{I}$ are
available ``free of charge'' whenever one performs closed testing for the
original $n$ elementary hypotheses $H_{i}.$ This follows from the fact that
``the closure of the closure is the closure;'' that is, that no new hypotheses
are generated when the set of $2^{n}-1$ intersection hypotheses is treated as
the set of elementary hypotheses. Hence, in GS's Figure 1, the significance of
$H_{\{2,3\}}$ can be stated with full FWER control over the set of $2^{3}-1=7$
hypotheses, and the conclusion $\tau(\{2,3\})\leq1$ follows immediately.
Again, GS provide a service in reminding statisticians (or in teaching those
who have not heard about it in the first place) of this nice feature of closure.

GS's paper also implicitly explains the following paradox: while closure is
based on composite hypotheses, it is not true that more powerful composite
tests lead to more powerful closure-based multiple tests.  When considering
only the elementary hypotheses, Bonferroni (or MaxT) types of composite tests,
which are usually thought to be the least powerful of the class of composite
testing methods (e.g., Nakagawa, \citeyear{2004Nakagawa}), tend to give higher power for
closure-based multiple tests (Romano, Shaikh and Wolf, \citeyear{2011Romano}).  However, when
the goal is to establish how many true effects there might be among a
collection of hypotheses, GS suggest indeed that more powerful composite tests
lead to more powerful multiple tests.

\begin{figure*}

\includegraphics{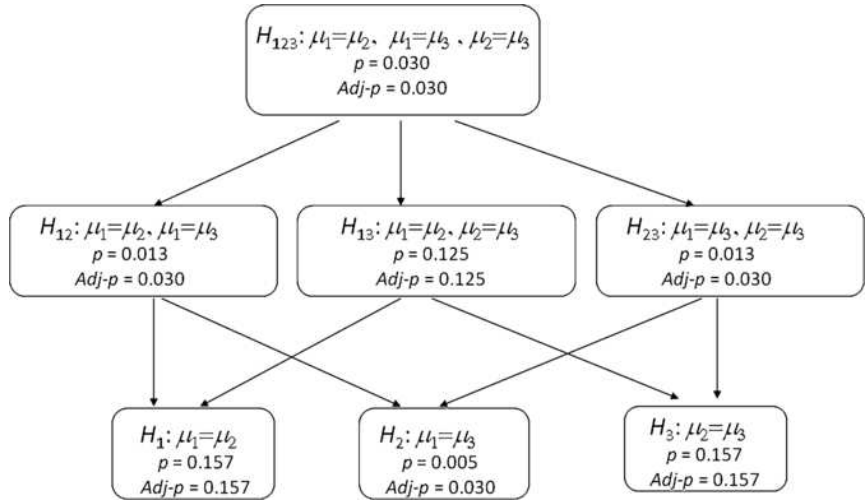}

  \caption{Closed testing with Fisher combination tests in a one-way ANOVA
setting, ignoring logical constraints.}\label{fig1}
\end{figure*}

The Fisher combination test is a useful choice of composite test, as noted by
GS.  But it is worth pointing out how bad this test can be compared to the
Bonferroni test, when both are used via closure for testing elementary
hypotheses. Consider analyzing a version (available from the author) of the
classic dataset reported by Golub at al. (\citeyear{1999Golub}), testing 7,129 genes for
association with either acute myeloid or acute lymphoblastic leukemia, using
7,129 two-sample $t$-tests.  The closed Fisher combination meth\-od is and has
been available in PROC MULTTEST of SAS/STAT with the $O(n^{2})$ shortcut since
release 8.1 of SAS in 2000; this software computes closure-based adjusted
$p$-values (defined below) to assess significance of elementary hypotheses.
Despite the fact that the Fisher combination test is liberal with correlated
data, the \textit{smallest} adjusted $p$-value using the closed Fisher
combination test is $1.000$ (rounded), hence none of the 7,129 tests are
significant at any reasonable nominal FWER level.  On the other hand, 37 of
the 7,129 genes have adjusted $p$-values less than the nominal 0.05 FWER level
when using closed Bonferroni (or Holm, \citeyear{1979Holm}) tests; the smallest adjusted
$p$-value is $1.7\times10^{-6}$ and is therefore extremely significant, even
after multiplicity adjustment.

I have some other comments/critiques about the paper that fall into the
following categories:\vadjust{\goodbreak}  (i) the~as\-sumption of free combinations and its
conse\-quences, (ii)~use of adjusted $p$-values rather than rigid nominal
thresholds, (iii)~computational shortcuts, and (iv) permutation testing.

\section{Additional Comments}

\subsection{Free Versus Restricted Combinations}

Implicit in GS's discussion of closure is that the elementary hypotheses obey
the \textit{free combinations} condition, which states that there are
$2^{n}-1$ distinct hypotheses in the closure.  Under restricted combinations
there are duplicates, and hence the set of intersections has many fewer
elements; by exploiting this fact one can obtain tighter confidence sets.
 For example, suppose $Y_{i}\sim N(\mu_{i},1)$, with $H_{1}\dvtx \mu_{1}=\mu_{2}$,
$H_{2}\dvtx \mu_{1}=\mu_{3}$ and $H_{3}\dvtx \mu_{2}=\mu_{3}$.  Then there are only four
elements in the closure rather than $2^{3}-1=7$, since $H_{\{1,2\}}%
=H_{\{1,3\}}=H_{\{2,3\}}=H_{\{1,2,3\}}$.  GS's method is valid but
conservative when all seven hypotheses are considered.

For example, suppose the data are $y_{1}=-2$, $y_{2}=0$ and $y_{3}=+2$,
yielding $z$-statistics $z_{1}=(-2-0)/\break 2^{1/2}=-1.414,$ $z_{2}=-2.828$ and
$z_{3}=-1.4141$, with corresponding two-sided $p$-values
$p_{1}=0.157299$,\break
$p_{2}=0.004678$ and $p_{3}=0.157299$.  The Fisher combinations statistics
are thus $c_{12}=c_{23}=-2\times\break \ln(0.157299\times 0.004678)=14.4291$, $c_{13}%
=7.3984$ and $c_{123}=18.1283$.  The chi-squared distribution cannot be used
to find $p$-values for these composite tests since the $Z$'s are not
independent, but under the null hypothesis, the vector of $Z$ statistics is
multivariate normal with\vspace*{2pt} mean vector $%
\biggl[{\fontsize{8.36}{8.36}\selectfont\matrix{
{0}\cr
{0}\cr
{0}
}}\biggr]
$ and covariance\vadjust{\goodbreak} matrix $%
\biggl[{\fontsize{8.36}{8.36}\selectfont\matrix{
1 & 0.5 & -0.5\cr
0.5 & 1 & 0.5\cr
-0.5 & 0.5 & 1
}}\biggr]
.$  Thus,\vspace*{3pt} the $p$-values can be obtained by simulating $Z$'s from this
distribution, computing the two-sided $p$-values, constructing the Fisher
combination statistics $C_{I}$, and counting how often the simulated $C_{I}$
exceeds the observed $c_{I}$. Figure~\ref{fig1} displays the results using these
$p$-values for each subset~$I$, as well as closure-based adjusted $p$-values.

Suppose that inference is considered for the set $\{H_{1},H_{3}\}$.  Here,
the confidence set for the number of true nulls is $\{0,1,2\}$, since $H_{13}$
is not rejected.  But the possibility that $\tau(\{1,3\})=2$ contradicts the
rejection of the global hypothesis $H_{123}$, and thus seems wrong.

\begin{figure}[b]

\includegraphics{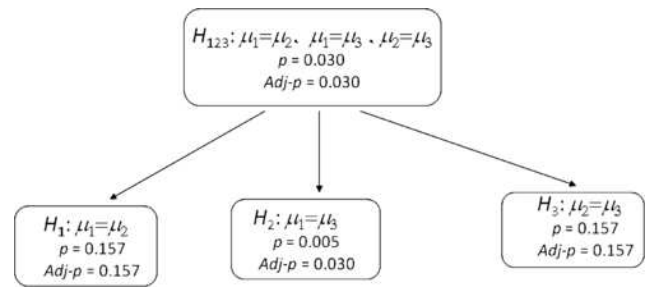}

  \caption{Closed testing with Fisher combination tests in a~one-way ANOVA
setting, incorporating logical constraints.}\label{fig2}
\end{figure}

Incorporating logical constraints, the graph is as shown in
Figure~\ref{fig2}.
Using logical constraints, the confidence set for $\tau(\{1,3\})$ is $\{0,1\}$
rather than $\{0,1,2\}$.

One can improve the power of closure-based consonant procedures as well by
utilizing logical constraints (Westfall and Tobias, \citeyear{2007Westfall}).

\subsection{Adjusted $p$-Values}

Adjusted $p$-values are simple and natural by-pro\-ducts of closure.  Let
$p_{I}$ be the local $p$-value for testing $H_{I}$. With closure, $H_{I}$ is
rejected only when~$H_{J}$ is~rejected for all $J\supseteq I$, or
equivalently, when\break $\max_{J\supseteq I}p_{J}\leq\alpha$, where $\alpha$ is the
nominal FWER.\break Hence $\max_{J\supseteq I}p_{J}$ is the adjusted $p$-value for
testing~$H_{I}$, and these are shown in my Figure~\ref{fig1}.

As GS note, exploratory inference should be \textit{mild}, \textit{flexible}
and \textit{post hoc}.  However, the use of a strict $0.05$ (or other)
nominal FWER threshold seems to violate the latter two of these
criteria.  For the same reasons that ordinary $p$-values are seen as more
natural and useful than the $0.05$-level determined ``accept/reject'' decision,
it is also more natural and useful to report an adjusted $p$-value along with
any claim about the number of true alternatives within a set of hypotheses.

For example, suppose my Figure~\ref{fig1} was from a case of free combinations, as with
GS's Figure~\ref{fig1}.  Then for the set $\{H_{1},H_{3}\}$, one cannot claim any
alternatives at the usual $0.05$ nominal FWER level, but one can conclude at
least one alternative at the nominal $0.13$ level.  The report could state
``For familywise significance levels as low as $0.125$, there is at least one
alternative among $\{H_{1},H_{3}\}$.''

In GS's discussion of Huang and Hsu's $n=4$ example where there are no
elementary significances, their conclusion is ``at least two out of the first
three hypotheses are false.''  After calculating the adjusted $p$-values for
these data, one can say ``at least two out of the first three hypotheses are
false (adjusted $p=0.038$).''  With other data, the conclusion might be that
``at least two out of the first three hypotheses are false (adjusted
$p=0.001$),'' which communicates quite different information, even though the
claimed number of alternatives is the same at the nominal FWER${}=0.05$ level.

Yet another benefit of adjusted $p$-values is that they offer a more realistic
assessment in the face of violated assumptions.  Assumptions are usually
wrong, and an adjusted $p$-value of $0.055$ might be more appropriately
reported as $0.041$ with a more correct analysis; conversely, $0.045$ might be
more appropriately reported as $0.053$.  Use of adjusted $p$-values rather
than fixed decisions better recognizes this fact, as savvy readers understand
that $p$-values are themselves approximations, and can use their own knowledge
or simulation studies to assess the accuracy of a ``$0.045$'' report.

A disadvantage of using adjusted $p$-values rather than
``accept/reject''
decisions\vadjust{\goodbreak} is that there are additional computations.  But this disadvantage
seems minor to me compared to problems with rigidly fixed nominal FWER levels.

\subsection{Computational Shortcuts}

The methodology GS espouse can be computatio\-nally prohibitive.  While closure
allows a simple~$O(n)$ shortcut in the case of the consonant Bonferroni--Holm
procedure, the GS methods will require something approaching~$O(2^{n})$
evaluations for most other cases of interest.   Shortcuts are available,
with less power as GS note.  Westfall and Tobias (\citeyear{2007Westfall}) use a tree-based
representation of the $2^{n}-1$ hypotheses, along with a branch-and-bound
algorithm for obtaining conservative, but computationally simpler analyses.
 These methods are available in a wide variety of SAS/STAT procedures as of
version 9.2 of SAS.

Oddly, GS do not mention Hommel's (\citeyear{1988Hommel}) $O(n^{2})$ closure shortcut when
using Simes' test; this shortcut is essentially identical to the one mentioned
by Zaykin et al. (\citeyear{2002Za}) for the truncated product (and by special case, Fisher
combination) test.

\subsection{Permutation Tests}

Permutation tests offer, under certain assumptions, exact rather than
approximate inference.  They also allow, in the case of binary data,
exceptionally higher power than corresponding methods based on continuous
data, by utilizing sparseness (Westfall, \citeyear{2010Westfall}).  In addition, tests that
assume independence require some correction for correlation structure, as
would be the case for the adverse event data of Table 3 of GS.  Hence,
permutation tests are useful for gaining power, as well as for obtaining valid
$p$-values.

Problems with permutation-based testing include computational difficulties and
hidden assumptions.  There is the obvious computational burden of either
enumerating or simulating the permutation distribution; doing this separately
for $O(2^{n})$ subsets is impossible, even for moderate $n$.  When the
``subset pivotality'' condition of Westfall and Young (\citeyear{1993Westfall}) is valid, one can
use a single global permutation distribution rather than $2^{n}-1$ separate
permutation distributions. The subset pivotality condition is valid for many
multivariate models, but fails for multiple comparisons with three or more
groups, since the global permutation distribution is not valid for making
pairwise comparisons involving two groups. If the subset pivotality condition
is satisfied, and if the (consonant) MinP tests are used, the computational\vadjust{\goodbreak}
burden is greatly reduced, making the Westfall--Young method feasible for
large-scale multiple testing applications.

One must also state their assumptions about the intersection hypotheses when
doing permutation-\break based analysis. When using permutation tests, the simplest
form of an elementwise null hypothesis is that the data are exchangeable
between groups.\break  However, the intersection of exchangeable elementwise
hypotheses does not imply joint exchangeability.  For example, consider the
two-group MANOVA with bivariate data.  If group one is bivariate normal with
mean vector $\mathbf{0}$ and identity covariance matrix, while group two
has the same mean vector\vspace*{2pt} but\break covariance  matrix $%
\left[{\fontsize{8.36}{8.36}\selectfont\matrix{
1.0 & 0.5\cr
0.5 & 1.0
}}\right]$,\vspace*{2pt} then the data in variable 1 are exchangeable between the groups
[specifically, i.i.d.~$N(0,1)$], the data in variable 2 are\break \mbox{exchangeable} between
the groups [also i.i.d. $N(0,1)$], but the two-dimensional vectors are not
exchangeable between the groups.  Thus, an assumption that mar\-ginal
exchangeability implies joint exchangeability is required when performing
permutation-\break based closed testing with multivariate multisample
data.\looseness=1

On the other hand, with consonant Bonferroni-based closed permutation
procedures, one can dispense with such assumptions.  These methods are
computationally simple, control the FWER for all sample sizes, and retain the
power advantage associated with permutation tests; details are given by
Westfall and Troendle (\citeyear{2008Westfall}).


\end{document}